\documentclass[final,5p,times,twocolumn]{elsarticle}

\usepackage{graphicx}

\usepackage{amssymb}

\journal{Nuclear Instruments and Methods A}

\begin{document}

\begin{frontmatter}



\title{Position resolution and efficiency measurements with large scale Thin Gap Chambers for the super LHC}

\author[A]{Nir Amram}
\author[A]{Gideon Bella}
\author[A]{Yan Benhammou} 
\author[B]{Marco A. Diaz}
\author[C]{Ehud Duchovni}
\author[A]{Erez Etzion}
\author[D]{Alon Hershenhorn}
\author[C]{Amit Klier}
\author[D]{Nachman Lupu}
\author[C]{Giora Mikenberg}
\author[C]{Dmitry Milstein}
\author[A]{Yonathan Munwes}
\author[E]{Osamu Sasaki}
\author[C]{Meir Shoa}
\author[C]{Vladimir Smakhtin\corref{cor1}}
\author[B]{Ulrich Volkmann}
\address[A]{Raymond and Beverly school of Physics and Astronomy, Tel Aviv University, Tel Aviv, Israel}
\address[B]{Pontifical Catholic University of Chile,Santiago,Chile}
\address[C]{Weizmann Institute of Science, Rehovot, Israel}
\address[D]{Technion Israel Institute of Technology, Haifa, Israel}
\address[E]{KEK, High Energy Accelerator Research Organization, Tsukuba, Japan}
\cortext[cor1]{Corresponding author. E-mail: Vladimir.Smakhtin@weizmann.ac.il (V.Smakhtin).}

\begin{abstract}
New developments in Thin Gap Chambers (TGC) detectors to provide fast trigger and high
precision muon tracking under sLHC conditions are presented. The modified detectors are shown
to stand a high total irradiation dose equivalent to 6~Coulomb/cm of wire, without showing
any deterioration in their performance. Two large (1.2 $\times$ 0.8~m$^2$) prototypes
containing four gaps, each gap providing pad, strips and wires readout, with a total
thickness of 50~mm, have been constructed. Their local spatial resolution has been
measured in a 100~GeV/c muon test beam at CERN. At perpendicular incidence
angle, single gap position resolution better than 60~$\mu$m has been obtained. For incidence angle 
of 20$^o$ resolution of less than 100~$\mu$m was achieved. 
TGC prototypes were also tested under a flux of 10$^5$~Hz/cm$^2$ of 5.5-6.5~MeV
neutrons, showing a high efficiency for cosmic muons detection.

\end{abstract}


\end{frontmatter}


\section{Introduction}
The expected background rates in the forward region of the ATLAS~\cite{ATLAS} Muon Spectrometer at the sLHC are expected to be 
higher by one order of magnitude compared to those at the LHC. Some of the present Muon Spectrometer components will fail to cope 
with these high rates and will have to be replaced. In particular, the expected rate of neutrons with energies above~100~keV 
may exceed 10$^5$~Hz/cm$^2$.

The higher rate of photons and minimum ionizing particles (MIPs) should be matched by a comparable increase in the rate capability 
of the various detector components. In addition, few MeV neutrons may give rise to very high signals due to the struggling of recoil
carbon nuclei from the gas and detector walls. The Thin Gap 
Chamber (TGC) detectors have been used successfully in the OPAL experiment 
at LEP~\cite{LEP} and are now a part of the ATLAS end-cap muon trigger system~\cite{muon}. 

\begin{figure}[hbt] 
\centering 
\includegraphics[width=0.34\textwidth,keepaspectratio]{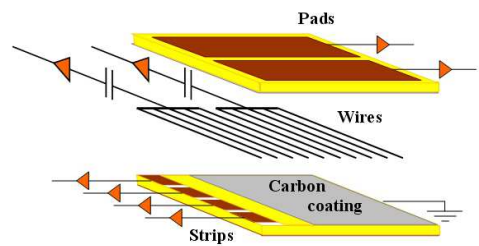}
\caption{Schematic view of the TGC.}
\label{Prototype}
\end{figure}

The TGC is a multiwire chamber with 50~$\mu$m diameter gold-plated tungsten wires, comprising the anode plane,
located between FR4 walls coated with resistive carbon that serves as the cathode.
The spacing between the wires is 1.8~mm and the anode-cathode spacing is 1.4~mm.
The operational gas is a mixture of 55$\%$ CO$_2$ and 45$\%$ n-pentane.
The carbon coating is transparent so that signals can be read out from strips or pads outside the gas volume.
A schematic view of the TGC with wire, strip and pad readout is shown in Fig.~\ref{Prototype}.

The present publication aims to show that a single layer TGC can provide spatial resolution better than 100~$\mu$m 
and hence to fulfill the ATLAS-sLHC requirement. 
This ability, coupled with the high rate capabilities and time resolution, makes the TGC technology a good candidate 
for triggering and tracking of high momentum muons at the sLHC.
The combination of a local trigger and good position resolution in the same gas gap would save precious space and 
simplify the structure of the future sLHC Muon Spectrometer. TGC trigger should provide fast signals, with an angular 
resolution of 1~mrad, within the present ATLAS timing restrictions of latency for the first level trigger.

\section{Aging and space charge effects}

\begin{figure}[hbt] 
\centering 
\includegraphics[width=0.2\textwidth,height=2.0in,keepaspectratio]{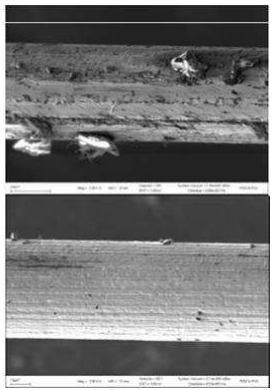}
\caption{A microscopical picture of the wires from the same detector, one of which was irradiated by the equivalent of 6~Coulomb/cm, while the other not.}
\label{Micro}
\end{figure}

\begin{figure}[hbt] 
\centering 
\includegraphics[width=0.25\textwidth,height=2.5in,keepaspectratio]{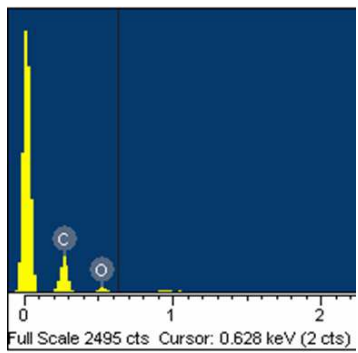}
\caption{Chemical spectrum of the irradiated wire deposits.}
\label{chemical_spectrum}
\end{figure}

\begin{figure}[hbt] 
\centering 
\includegraphics[width=0.3\textwidth,height=2.5in,keepaspectratio]{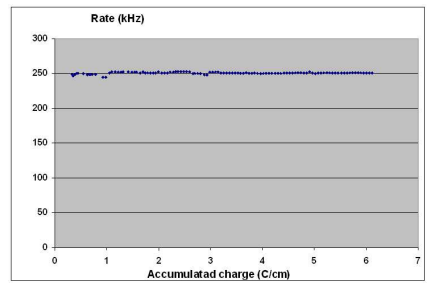}
\caption{TGC count rate vs. accumulated charge.}
\label{rate}
\end{figure}

The ability to withstand long term high radiation without showing aging effects is a prerequisite for any proposed 
device at the sLHC. Another concern is the deterioration of performance due to space charge effects. The detectors must, therefore, 
be radiation tolerant and should have inherent weak rate dependence. 
Aging effects were studied by irradiating a small (10$\times$10~cm$^2$) TGC detector with a $^{90}$Sr beta source (with electron energies of 0.546 
and 2.283~MeV) during a five months period. The irradiation was followed by a microscopic and chemical analysis of the wire 
deposits. The microscope picture in Fig.~\ref{Micro} shows that the deposits are widely spaced, which leads to negligible changes 
in the electric field between anodes and cathodes. The chemical spectrum in Fig.~\ref{chemical_spectrum} shows that the 
deposits consist  mainly of carbon and oxygen, which are clearly related to the gas components and not to impurities in the 
gas system. The TGC anode count rate as a function of the accumulated charge is shown in Fig.~\ref{rate}. No rate decrease has been 
observed. The detector showed no deterioration in its response to electrons after being irradiated by the equivalent of 
6~Coulomb/cm (approximately 100 years of sLHC).
Space charge effect measurements are described in detail in~\cite{Fukui}. No space charge accumulation was observed up to 
100~kHz/cm$^2$ signal hit rate.

\section{Design and construction of the TGC Large Prototype}
Two large prototypes, that can be assembled into a full size detector, 1.2 $\times$ 0.8~m$^2$ each, were constructed. 
Each prototype includes four layers of TGCs (their layout is shown schematically in Fig.~\ref{Prototype}) and fit in a total 
thickness of  50~mm. Each layer contains a series of pads for local trigger coincidence and 
strips, perpendicular to the wires, for high precision position measurement as well as local precision trigger elements.
The wires are used for a second coordinate measurement.
The large prototype parameters are shown in Table~\ref{prototype_params}.

\begin{table}
\begin{tabular}{|l|l|}
\hline
\multicolumn{2}{|c|}{Large prototype parameters} \\
\hline
Strip-carbon gap & 0.1 mm \\
Strip pitch & 3.2 mm \\
Inter-strip gap & 0.3 mm \\
Wire length in 4 layers & 0.4, 0.5, 0.6, 0.7 m \\
Number of wires ganged together & 5 (9~mm granularity) \\
Strip length & 1.2 m \\
Pad size & 40$\times$10 $\mathrm{cm}^{2}$ \\
Carbon surface resistivity & 50~k$\Omega$/square \\
HV blocking capacitance & 470~pF \\
\hline
\multicolumn{2}{|c|}{Readout electronics parameters} \\
\hline 
Preamplifier gain & 0.8V/pC \\
Integration time & 16 ns \\
Main amplifier gain & 7 \\
Equivalent noise charge & 7500 electrons \\
 & at CD=150 pF \\
\hline
\end{tabular}
\caption{Large prototype parameters.}
\label{prototype_params}
\end{table}

\section{Position resolution test beam at CERN}
The position resolution of one of the large TGC prototypes 
was measured using 100~GeV/c muons from the SPS-H8 test beam 
at CERN in June 2009. Previous CERN pion test beam results with smaller prototypes are
described in~\cite{NIMarticle}. The main goal of the test was to determine the
position resolution in each of the layers using analog and fast digital readout, as 
well as its dependence on the muon incidence angle. Each detector was equipped with 16 strip 
analog and digital readout channels of similar type as those used in the ATLAS TGC~\cite{Sasaki}.

\begin{figure}[hbt] 
\centering 
\includegraphics[width=0.3\textwidth,height=3.0in,keepaspectratio]{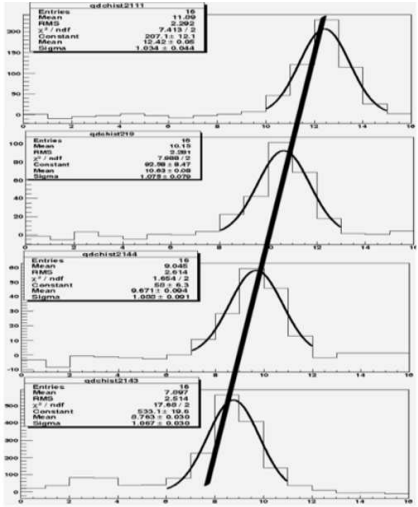}
\caption{Typical TGC quadruplet event as seen in the ADC counts histograms.}
\label{event}
\end{figure}

The external trigger was provided by a coincidence of two plastic scintillators.
The position resolution is directly related to the profile of induced charge on the strips and on the accuracy of charge measurement. 
The actual charge on each of the strips was measured using two 32-channel, 12-bit resolution charge integrating ADC modules CAEN V792. 
The four ADC count distributions for a typical event in the 4-layer prototype are shown in Fig.~\ref{event}.
The four track hit positions were
determined by a Gaussian fit. For each layer, the expected position was determined by the extrapolation 
of the positions in the other three layers. The difference between the measured position and the expected 
position was defined as the residual.

\begin{figure}[hbt] 
\centering 
\includegraphics[width=0.34\textwidth,keepaspectratio]{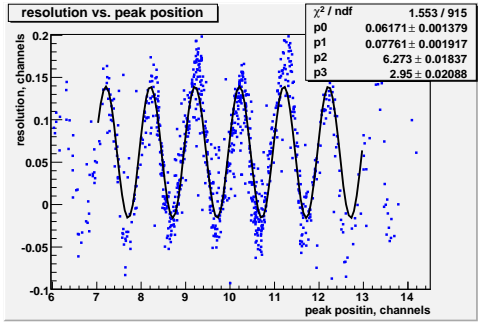}
\caption{Residual vs. hit position with sinusoidal fit.}
\label{nonlin}
\end{figure}

\begin{figure}[hbt] 
\centering 
\includegraphics[width=0.34\textwidth,keepaspectratio]{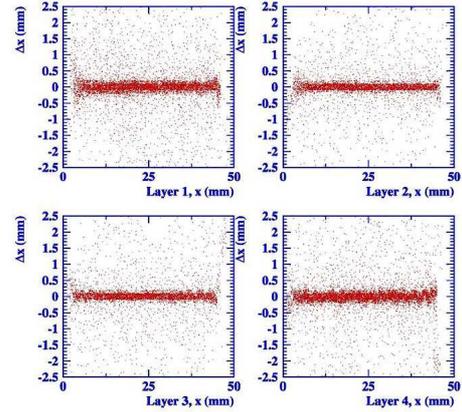}
\caption{Residual vs. hit position for the four TGC protorypes after differential non-linearity correction.}
\label{resolution}
\end{figure}

Due to the periodic strip structure, the deviation of the expected hit position from the measured one depends on 
the hit position. This dependence is clearly seen in Fig.~\ref{nonlin}. 
A sinusoidal fit was applied to correct for this differential non-linearity effect. The final deviation was calculated 
from the fit curve. The residual distributions for each of the four detectors vs. hit position after this correction is shown 
in Fig.~\ref{resolution}. The width of this distribution, $\sigma_{residual}$, is defined as the local resolution for 
each detector.

The resolution is also affected by the incidence angle $\phi$, since this angle determines the track projection 
on the anode wire, as explained in~\cite{NIMarticle}. 

\begin{figure}[t] 
\centering 
\includegraphics[width=0.35\textwidth,height=3.5in,keepaspectratio]{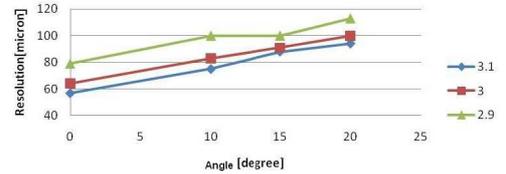}
\caption{Local resolution vs. incidence angle for different HV values.}
\label{resolution_angle1}
\end{figure}

The angular dependence of the resolution was studied by rotating the chambers with respect to the beam axis. 
A set of measurements with different anode high voltage (HV) values were performed for incidence angles of: 0, 5, 10, 15, and 
20 degrees. 
three HV values were used: 2.9, 3.0, and 3.1~kV.
The local spatial resolution vs. incidence angle for different HV values is shown in Fig.~\ref{resolution_angle1}.
One can see that the resolution deteriorates as the angle increases and improves with higher HV. Single gap resolution of  
better than 100~$\mu$m  is achieved for incidence angles of up to 20$^o$ and HV value of 3.0~kV.
These values of resolution meet the sLHC requirements.

\begin{figure}[t] 
\centering 
\includegraphics[width=0.35\textwidth,height=3.5in,keepaspectratio]{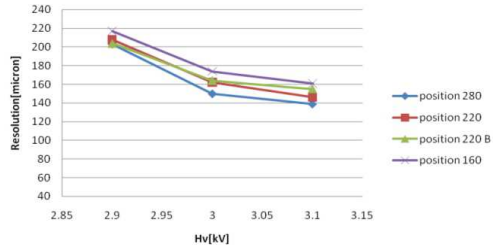}
\caption{Digital resolution vs. HV for four different geometrical TGC positions at zero incidence angle.}
\label{digital_res}
\end{figure}

In order to confirm the suitability of the TGC as a trigger device, we also measured the spatial resolution of the TGC with digital
signals from the strips using the time over threshold information for each channel. The time was measured with a 
VME 32CH TMC TEG3 KEK module. Determination of the hit position and the resolution measuring were done the same way 
as for the analog readout. The obtained digital resolution values vs. HV for four different geometrical TGC positions 
(in order to avoid the geometrical systematic) at perpendicular incidence are shown in Fig.~\ref{digital_res}. At 3.0~kV the 
average digital resolution is 160~$\mu$m which easily meets the level one trigger angular resolution requirement 
of 1~mrad for a 200~mm thick detector.

\section{Neutron beam test at Demokritos}
\begin{figure}[hbt] 
\centering 
\includegraphics[width=0.35\textwidth,height=3.5in,keepaspectratio]{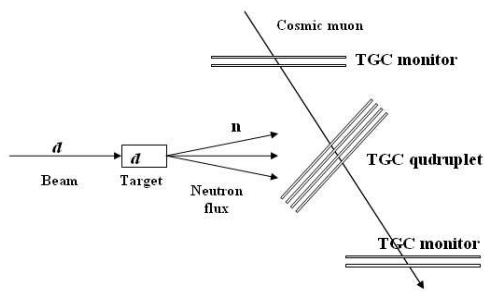}
\caption{Experimental setup at Demokritos.}
\label{test_beam_neutron}
\end{figure}

\begin{figure}[t] 
\centering 
\includegraphics[width=0.35\textwidth,height=3.5in,keepaspectratio]{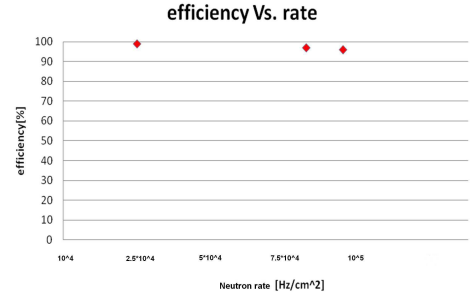}
\caption{Muon detection efficiency vs. detected neutron rate.}
\label{efficiency_neutron}
\end{figure}

Muon detecting capabilities of the TGC were also studied when the detector was exposed to a high flux of neutrons.
The test was carried out at Demokritos, Greece in October 2009.
The facility at Demokritos consists of a 5.5~MV TANDEM T11/25 accelerator, which uses Van de Graaff
electrostatic technique with high voltages between 0.4 and 5.5~MV. For the TGC test, neutrons with an energy 5.5-6.5~MeV 
were produced via the $^2$H($^2$H,n)$^3$He reaction. The detailed description of the Demokritos facility can be found 
in~\cite{Demokritos}.
Neutron flux measurement was performed via the $^{191}$Ir(n,2n)$^{190}$Ir activation reaction~\cite{Iridium}.
A schematic view of the experimental setup is shown in Fig.~\ref{test_beam_neutron}. 
The trigger for cosmic ray muons was 
provided by the triple coincidence of two pairs of "monitors" -- small TGC prototypes
(16$\times$12~cm$^2$) with 3 and 4~mm strips on both sides.
The hit position was obtrained by four other identical small prototypes (the "quadruplet") 
placed between the monitor pairs, as shown in the schematic.
The same type of front-end and readout electronics as in the CERN test was used. The procedure of determining the efficiency 
and resolution was identical to the one described in the previous section, i.e. by fitting a track using three TGC layers and 
comparing the predicted and measured positions on the fourth.
A hit was considered good if the measured position was less than 10~mm away from the predicted one.
Muon detection efficiencies vs. neutron rates are shown in Fig.~\ref{efficiency_neutron}. The efficiency deterioration
at high rate is not significant. Although there was a concern that neutrons may give rise to large signals, producing
HV breakdown (sparks), no such sparks were observed during the five-day period of the test.

\section{$^{60}$Co test at Soreq}
The TGC cosmic muon efficiency was also tested under a high flux of gamma photons from a 47~Ci $^{60}$CO source 
at the Soreq Nuclear Research Center, Israel.
While the experimental setup was very similar to the setup at Demokritos (using a large prototype instead of the small quadruplet),
no efficiency deterioration was observed under a gamma flux
of 2$\cdot$10$^6$~Hz/cm$^2$, which is 100 times the LHC gamma flux. Efficiency measurements under higher fluxes up 
to 10$^7$~Hz/cm$^2$ will also be performed.

\section{Conclusions}
Two large prototypes
were constructed and tested in the SPS-H8 test beam 
at CERN and in a neutron test beam at Demokritos. They show good rate capabilities 
over large areas, good single gap position resolution (better than 60~$\mu$m for minimum ionizing 
particles at perpendicular incidence angle), fast response, and 
the option to combine trigger and tracking in the same gas gap, all at a 
reasonable cost. The TGC detectors showed almost no decrease in muon detection efficiency
while operating under 10$^5$~Hz/cm$^2$ flux of 5.5-6.5~MeV neutrons as well as
under 2$\cdot$10$^6$~Hz/cm$^2$ flux of gammas, also no HV sparks were observed.
Detector irradiation to a total accumulated charge of 6~Coulomb/cm of wire shows no deterioration
in its performance.

\section{Acknowledgments}
The authors would like to thank the members of the ATLAS TGC group, 
especially R. Alon, M. Ben  Moshe and B. Yankovsky, our colleagues from 
the Soreq Nuclear Research Center, in paticular I. Hershko, D. Breitman, and 
our colleagues from National Center for Scientific Research ``Demokritos'', Greece, especially G. Fanourakis and T.Geralis, 
as well as T. Alexopoulos and M. Dris from the National Technical University of Athens 
for their support of this work. This work was partly supported by the Benoziyo Center 
for High Energy Physics, The Israeli Science Foundation (ISF) and the Minerva Foundation.

\end{document}